\documentstyle[preprint,aps]{revtex}
\begin{document}
\draft

\title{Amplitude ambiguities in pseudoscalar meson photoproduction}
\author{Greg Keaton and Ron Workman}
\address{Department of Physics, Virginia Polytechnic Institute and State
University, Blacksburg, VA~24061}

\date{\today}
\maketitle

\begin{abstract}

We consider the problem of determining amplitudes from observables for the
case of pseudoscalar meson photoproduction.
We find a number of surprisingly simple constraints which
give necessary conditions for a complete
set of measurements. These results contradict one
of the selection rules derived previously.

\end{abstract}

\eject

\narrowtext

Experiments conducted at CEBAF will soon yield a flood of new and precise
data for the photo- and electroproduction of mesons. This has motivated
a renewed examination of the observables required for the determination
of underlying amplitudes. The electroproduction of pseudoscalar mesons
has been studied by Dmitrasinovic, Donnelly and Gross\cite{Dmit} and,
more recently, the photoproduction of vector mesons has been studied
by Tabakin and co-workers\cite{Tab}. This type of analysis was
applied to the photoproduction of pseudoscalar mesons by
a number of groups. The work of Barker, Donnachie and Storrow\cite{Bark}
is generally quoted as the standard reference.

In Ref.\cite{Bark}, the requirements for a complete set of measurements
were studied within the transversity representation. This representation
is particularly useful as measurements of the cross section and single
polarization observables directly determine the magnitudes of the 4
independent amplitudes. It then remains to determine relative phases from
double polarization measurements. In Ref.\cite{Bark} it was claimed that
5 double polarization measurements are required in order to resolve all
ambiguities (apart from an overall phase) in the transversity
amplitudes. In choosing the 5 double polarization measurements one had
only to insure that fewer than 4 were taken from any one set of beam-target
(BT), beam-recoil (BR), or target-recoil (TR) observables.
Only 3 double polarization measurements (not all from the same set)
were found necessary
to determine the amplitudes up to "quadrant ambiguities".

As analyses are generally performed using helicity amplitudes, we reexamined
the question of complete experiments within this basis. (In order to avoid
confusion, we retain the naming scheme of Ref.\cite{Bark}.) The
4 independent helicity amplitudes are denoted $S_1$, $S_2$ (single spin-flip),
$N$ (no spin-flip), and $D$ (double spin-flip). Apart from overall factors,
the relations between amplitudes and observables are given in Table~I.
If one examines the set of cross section and single polarization measurements,
a number of ambiguities are evident. The first (trivial) ambiguity results
from the freedom to alter the overall phase of these amplitudes.
It is not hard to
find 3 more ambiguity relations associated with the set of cross section and
single polarization observables.

\eject

\centerline{Ambiguity I}
\begin{equation}
S_1 \leftrightarrow S_2  \;\; {\rm and} \;\;  N \leftrightarrow -D
\end{equation}
\centerline{Ambiguity II}
\begin{eqnarray}
S_1 & \to N \cr
N   & \to -S_1 \cr
S_2 & \to -D \cr
D   & \to S_2
\end{eqnarray}
\centerline{Ambiguity III}
\begin{eqnarray}
S_1 & \to D \cr
D   & \to -S_1 \cr
N   & \to S_2 \cr
S_2 & \to -N
\end{eqnarray}
If the operations indicated above are carried out, the first four observables
listed in Table~I are unchanged. The remaining double polarization
observables are either
unchanged, or changed by at most a sign.

The operations I, II and III are associated with ambiguities in the
TR, BR, and BT observables respectively\cite{compose}.
These discrete symmetries are in fact special cases of three continuous
symmetries\cite{Dean}
which correspond to changing the three angles (left unspecified
by single polarization measurements) between the four transversity
amplitudes ($b_1$, $b_2$, $b_3$, and $b_4$).
In order to determine these angles, up to discrete quadrant ambiguities,
we require a set of three double-polarization measurements which is not
invariant under operations I, II, or III. Therefore, no more than two
measurements can be of the same type (see Table~I). This is identical to
the requirement stated in Ref.\cite{Bark}.

If we were only able to produce necessary conditions which agreed with the
results of Ref.\cite{Bark}, this approach would have limited usefulness.
However, the determination of a set of amplitudes which resolves {\it all}
ambiguities, apart from the trivial one, is a much more difficult problem.
Here the utility of this method becomes clear.
In order to explore this problem, and test the conclusion of Ref.\cite{Bark},
we generate one additional ambiguity involving
complex conjugation\cite{trans}.
\centerline{Ambiguity IV}
\begin{eqnarray}
S_1 & \to - S_1^* \cr
S_2 & \to - S_2^* \cr
N  & \to N^* \cr
D & \to D^*
\end{eqnarray}
If the statements\cite{Bark} regarding discrete ambiguities
were correct, then by
choosing no more than 2 observables from each double polarization set,
we would resolve this ambiguity as well.
However, it is easy to see that certain choices will
leave ambiguity IV unresolved. In fact, we can measure the 6 observables
$H$, $E$, $O_x$, $C_z$, $T_x$, and $L_z$ without resolving this ambiguity.
Further constraints on observable choices can be generated by composing the
transformation given in set IV with sets I, II and III. The resulting
additions to Table~I are easily found by multiplying elements from the
corresponding columns.

In Ref.\cite{Bark}, the resolution of quadrant ambiguities was demonstrated
for measurements of $G$, $F$, $E$, $L_x$, and one other observable not
from the BT set. This set does indeed resolve all of the ambiguities listed
in Table~I. As the authors point out, the enumeration of all possibilities
is ``exceedingly tedious'' (the `no four from any set' criterion allows
768 possible combinations). It is easy to see how the constraint given by
Ambiguity IV could be missed, since it involves measurements from all three
double polarization sets.

In summary, the examination of ambiguity relations provides a simple
and useful check of proposed complete sets of experiments. We have found
that the rules for choosing observables are more complicated than those
given in Ref.\cite{Bark}.
Note that some measurements of $G$, $H$, and $O_x$ exist, and these
measurements resolve all the ambiguities listed above.
Unfortunately, while {\it necessary} conditions are relatively
easy to generate, the proof of {\it sufficient} conditions is
still difficult. We are continuing to work on this aspect of the problem.

This work was supported in part by the U.S. Department of
Energy Grants DE-FG05-88ER40454 and DE-FG05-95ER40709A.

\eject

\newpage

\begin{table}
\caption{Result of ambiguity relations applied to observables.
Overall factors have been removed in the relations between amplitudes and
observables. The observables are either invariant (+) or change sign ($-$)
under these operations.}
\label{tbl3}
\vskip .2cm
\begin{tabular}{lcccccc}
{\bf Observable} & Type & Helicity Rep. & ( I ) & ( II ) & ( III ) & ( IV )\\
\hspace*{0.15in} &  &  &  &  &  &                                   \\
$\sigma (\theta)$& & $|N|^2 +|S_1|^2 + |S_2|^2 + |D|^2$ & + & + & +&+ \\
$\Sigma$     & S   & 2 Re $(S_1^* S_2 - N   D^*)$       & + & + & +&+ \\
$T$             &  & 2 Im $(S_1   N^* - S_2 D^*)$       & + & + & +&+ \\
$P$             &  & 2 Im $(S_2   N^* - S_1 D^*)$       & + & + & +&+ \\
\hspace*{0.15in} & &  &  &  &  &                                 \\
$G$             &  & $-2$ Im $(S_1  S_2^* + N D^*)$     & $-$&$-$&+&$-$\\
$H$          & BT  & $-2$ Im $(S_1  D^*  + S_2 N^*)$    & $-$&$-$&+&+\\
$E$             &  & $|S_2|^2 -|S_1|^2 - |D|^2 + |N|^2$ & $-$&$-$&+&+\\
$F$             & & 2 Re $(S_2  D^* + S_1 N^*)$        & $-$&$-$&+&$-$\\
\hspace*{0.15in} & &  &  &  &  &                                \\
$O_x$           &  & $-2$ Im $(S_2   D^* + S_1 N^*)$    & $-$&+&$-$&+\\
$O_z$       & BR   & $-2$ Im $(S_2 S_1^* + N   D^*)$    & $-$&+&$-$&$-$\\
$C_x$            & & $-2$ Re $(S_2  N^* + S_1 D^*)$   & $-$&+&$-$&$-$\\
$C_z$          &  & $|S_2|^2 -|S_1|^2 - |N|^2 + |D|^2$  & $-$&+&$-$&+\\
\hspace*{0.15in} & &  &  &  &  &                               \\
$T_x$           &  & 2 Re $(S_1 S_2^* +  N   D^*)$    & +&$-$&$-$&+\\
$T_z$      & TR    & 2 Re $(S_1 N^* - S_2  D^*)$      & +&$-$&$-$&$-$ \\
$L_x$           &  & 2 Re $(S_2  N^* - S_1 D^*)$      & +&$-$&$-$&$-$ \\
$L_z$         &  & $|S_1|^2 +|S_2|^2 - |N|^2 - |D|^2$  & +&$-$&$-$&+\\
\hspace*{0.15in} & &  &  &  &  &                             \\
\end{tabular}
\end{table}

\begin{thebibliography}{99}
\bibitem{compose} Note that Ambiguity I can be obtained from the composition
of Ambiguities II and III.
\bibitem{trans} In the transversity representation, we can apply the
               transformation: ($b_1$, $b_2$, $b_3$, $b_4$) $\to$
                ($b_1^*$, $b_2^*$, $b_3^*$, $b_4^*$).
\end{thebibliography}
\end{document}